\documentclass[sigconf, manuscript]{acmart}

\setcopyright{none}
\usepackage{subcaption}
\usepackage{graphics}
\usepackage{caption}
\usepackage{float}
\usepackage{makecell}
\usepackage{tabularx}
\usepackage{natbib}

\usepackage{booktabs}
\usepackage{subcaption}  

\usepackage{geometry}
\graphicspath{{figures/}}

\AtBeginDocument{%
	}

\begin{document}
	
\title{Wearable Music2Emotion : Assessing Emotions Induced by AI-Generated Music through Portable EEG-fNIRS Fusion}


\author{Sha Zhao}
\authornote{These authors contributed equally to this work.}
\affiliation{
	\institution{Zhejiang University}
	\city{Hangzhou}
	\country{China}
}
\email{szhao@zju.edu.cn}

\author{Song Yi}
\authornotemark[1]
\affiliation{
	\institution{Zhejiang University}
	\city{Hangzhou}
	\country{China}	
}
\email{syi@zju.edu.cn}

\author{Yangxuan Zhou}
\affiliation{
	\institution{Zhejiang University}
	\city{Hangzhou}
	\country{China}
}
\email{zyangxuan@zju.edu.cn}

\author{Jiadong Pan}
\affiliation{
	\institution{Hangzhou RongNao  Technology Co., Ltd}
	\city{Hangzhou}
	\country{China}
}
\email{jdpan@zju.edu.cn}

\author{Jiquan Wang}
\affiliation{
	\institution{Zhejiang University}
	\city{Hangzhou}
	\country{China}
}
\email{wangjiquan@zju.edu.cn}

\author{Jie Xia}
\affiliation{
	\institution{Hangzhou RongNao  Technology Co., Ltd}
	\city{Hangzhou}
	\country{China}
}
\email{JieXia@zju.edu.cn}

\author{Shijian Li}
\affiliation{
	\institution{Zhejiang University}
	\city{Hangzhou}
	\country{China}
}
\email{shijianli@zju.edu.cn}

\author{Shurong Dong}
\affiliation{
	\institution{Zhejiang University}
	\city{Hangzhou}
	\country{China}
}
\email{dongshurong@zju.edu.cn}

\author{Gang Pan}
\authornote{Corresponding author: Gang Pan.}
\affiliation{
	\institution{Zhejiang University}
	\city{Hangzhou}
	\country{China}
}
\email{gpan@zju.edu.cn}


\renewcommand{\shortauthors}{Sha Zhao et al.}


\pagestyle{fancy}
\fancyhead{}
\fancyfoot{}

\fancypagestyle{firstpagestyle}{
	\fancyfoot{}
	\fancyhead[L]{\textbf{Preprint version. Accepted by ACM MM 2025.}}
	\renewcommand{\headrulewidth}{0.3pt}
}

\fancyhead[L]{\textbf{Pre-print version. Accepted by ACM MM 2025.}}
\renewcommand{\headrulewidth}{0.3pt}

	\begin{abstract}
		Emotions critically influence mental health, driving interest in music-based affective computing via neurophysiological signals with Brain-computer Interface techniques. While prior studies leverage music's accessibility for emotion induction, three key limitations persist: \textbf{(1) Stimulus Constraints}: Music stimuli are confined to small corpora due to copyright and curation costs, with selection biases from heuristic emotion-music mappings that ignore individual affective profiles. \textbf{(2) Modality Specificity}: Overreliance on unimodal neural data (e.g., EEG) ignores complementary insights from cross-modal signal fusion.\textbf{ (3) Portability Limitation}: Cumbersome setups (e.g., 64+ channel gel-based EEG caps) hinder real-world applicability due to procedural complexity and portability barriers. To address these limitations, we propose MEEtBrain, a portable and multimodal framework for emotion analysis (valence/arousal), integrating AI-generated music stimuli with synchronized EEG-fNIRS acquisition via a wireless headband. By MEEtBrain, the music stimuli can be automatically generated by AI on a large scale, eliminating subjective selection biases while ensuring music diversity. We use our developed portable device that is designed in a lightweight headband-style and uses dry electrodes, to simultaneously collect EEG and fNIRS recordings. A 14-hour dataset from 20 participants was collected in the first recruitment to validate the framework's efficacy, with AI-generated music eliciting target emotions (valence/arousal). We are actively expanding our multimodal dataset (44 participants in the latest dataset) and make it publicly available to promote further research and practical applications. \textbf{The dataset is available at} \url{https://zju-bmi-lab.github.io/ZBra}.
	\end{abstract}

	
	
	
	
	
	\maketitle
	
	\section{Introduction}
	
	Emotions significantly impact human health and are linked to mental health disorders, such as depression, anxiety, attention deficit hyperactivity disorder, and internet addiction \cite{desteno2013affective, north2004uses,kubzansky2000going}. This underscores the necessity of accurate emotion recognition and effective regulation strategies in disease prevention and therapeutic management. Music facilitates emotional regulation, supported by its prevalent  application in daily stress reduction and mood enhancement. Compared to other media (e.g., videos), music's high accessibility makes it a more practical and scalable tool for emotional regulation. Prior research has examined music-induced emotions and their interplay \cite{atkinson2016improving, bhosale2022calibration}, yet current paradigms primarily rely on subjective self-reports to evaluate emotional responses. Such approaches are inherently limited by introspective biases (e.g., retrospective inaccuracies) and inter-subject variability in affective labeling \cite{moore2013systematic, north2004uses, north2008social}. Thus, establishing an objective mapping between affective states and music patterns is critical, as this systematic approach is imperative to advance toward precision music-based emotion regulation.
	
	\textbf{Emotions are closely tied to neural dynamics of the brain.} For instance, elevated affective states (e.g., joy) correlate with heightened cortical activation, whereas depressive states associate with reduced neural oscillatory activity. Such neural dynamics can be captured via brain-computer interface (BCI) systems, which decode neural signatures and enable direct interaction with external environments. Consequently, non-invasive BCI modalities (e.g., EEG, electroencephalography) offer a valid framework for objective emotion research due to their intrinsic safety and ease of use. Emerging studies have leveraged these techniques to discriminate affective states through neural signals \cite{koelsch2010towards, koelsch2014brain, kim2008emotion, wang2024cbramod, zhoubrainuicl, wang2025m, wang2024diffmdd}. Notably, recent advances have targeted music-induced emotion recognition and regulation through brain signals, demonstrating efficacy \cite{kaneshiro2016naturalistic, calcagno2025eeg, ehrlich2019closed}. 
	
	\textbf{Nevertheless, critical limitations persist on the stimuli music, brain signal modality, and device portability}. \textbf{(1) Musical stimuli limitation}: Existing studies predominantly employ small-scale music corpora (e.g., 6 \cite{korhonen2006modeling}, 16 \cite{lin2010eeg} 
	and 20 \cite {thoma2006regulation}), constrained by copyright barriers and curation costs \cite{moore2013systematic}. This could cause statistical bias or even risking model overfitting during emotion recognition. Moreover, the music stimuli are usually chosen subjectively based on experimenters’ heuristic assumptions about emotion-music mappings, neglecting idiosyncratic affective profiles across individuals. It could reduce the model robustness and generalizability \cite{eerola2012review, moore2013systematic}. Besides, widely recognized musical pieces activate pre-existing autobiographical associations, thereby negatively affecting intended emotion induction through extraneous cognitive processing (e.g., memory retrieval) rather than acoustically-driven affective responses\cite{lin2010eeg, moore2013systematic, gomez2007relationships}. \textbf{(2) Brain signal modality constraint}: Most of affective computing paradigms rely on unimodal neural signal (e.g., EEG-only) \cite{calcagno2025eeg}. While EEG excels in temporal resolution and remains the dominant modality for non-invasive affective computing \cite{lin2010eeg, thammasan2016application}, its spatial specificity is intrinsically constrained by scalp-level signal attenuation. Multimodal fusion (e.g., EEG-fNIRS, functional Near Infrared Spectroscopy) could mitigate this by providing complementary clues for analyzing emotions. For instance, fNIRS capture distinct patterns of hemodynamic responses \cite{ding2024understanding}, and performs well in emotion recognition \cite{qiu2022multi}. \textbf{(3) Portability limitation}: While music’s intrinsic accessibility facilitates emotion induction in a convenient fashion, existing paradigms overlook another critical role to promote the music-induced emotion: the device portability. Cumbersome setups (e.g., 64+ channel EEG caps requiring conductive gel) demand specialized infrastructure and operator expertise, increasing costs and restricting scalability \cite{atkinson2016improving, bhosale2022calibration}. Multimodal monitoring (e.g., EEG-EMG, electromyogram-ECG, electrocardiogram) necessitates more than 30 minutes for setup\cite{ehrlich2019closed}. These operational bottlenecks fundamentally constrain real-world applicability. 
	
	To address the aforementioned limitations, we propose a portable and multimodal framework for analyzing emotions induced by AI-generated music and collecting EEG and fNIRS through our portable headband. For the first challenge, we use music generated by AIGC (AI-Generated Content) to induce emotions instead of selecting well-known music pieces, overcoming the limitation in the scale, diversity, subjective selection and bias emotion responses. We first design prompt templates based on the Valence-Arousal model \cite{russell1980circumplex}, and then the prompt sentences created by the templates are input to automatically generate music clips. For the second and third challenges, we use our developed portable device to simultaneously collect EEG and fNIRS recordings both of which are used for emotion recognition. Our device is designed in a lightweight headband-style and uses dry electrodes, making it easy to wear and user-friendly. We design data collection paradigm and acquire multimodal brain signals in the real-world. On the basis of the dataset, the generated music clips are verified and the emotion-inducing effects are evaluated, indicating the generated music effectively evokes the intended emotions. Our contributions are as follows:

	\begin{itemize} 
		\item We propose a portable and multi-modal framework for systematic acquisition of brain signals induced by AI-generated music, through a wireless portable headband. Evaluated on our collected real-world dataset, this framework overcomes the limitations in music stimuli, signal modality and portability, promoting the real-world applicability and accessibility of emotion regulation.
		
		\item The music stimuli are automatically generated by AIGC techniques, which allow for creation on a large scale without relying on subjective selection. Verified on our real-world dataset, the generated music can effectively evokes emotions, addressing the issues in terms of the small scale, limited diversity and bias emotion responses.
		
		\item The acquisition device is portable and multimodal integrated, easy to wear and simultaneously recording EEG and fNIRS for emotion analysis. We collected a real-world dataset of EEG and fNIRS from 20 individuals in the first recruitment, with a total duration of about 14 hours. We are actively expanding our multimodal dataset (44 participants in the latest dataset) and make it publicly available to promote further research and practical applications.
	\end{itemize}

	\section{Related Work}
	\vspace{0.2in}	
	\subsection{Music-Induced Emotions and Physiological Responses}
	\vspace{0.15in}
	
	The studies of music-induced emotions have been guided by two dominant theoretical models: the discrete emotion model and the dimensional/continuous emotion model, which includes \textit{Valence}-\textit{Arousal} bipolar coordinate system proposed by \citet{russell1980circumplex}.
	The discrete emotion model posits that emotions fall into fundamental categories such as happiness, misery, and anger, each of which is biologically "hardwired" and characterized by distinct and separate physiological, neural, and expressive markers \cite{colombetti2009affect, lin2010eeg}.
	In contrast, the dimensional/continuous emotion model represents emotions along continuous axes, such as \textit{Valence} (pleasantness-unpleasantness) and \textit{Arousal} (activation-deactivation) \cite{hsu2017automatic, koelstra2011deap}.

	\begin{figure*}
		\includegraphics[width=\textwidth]{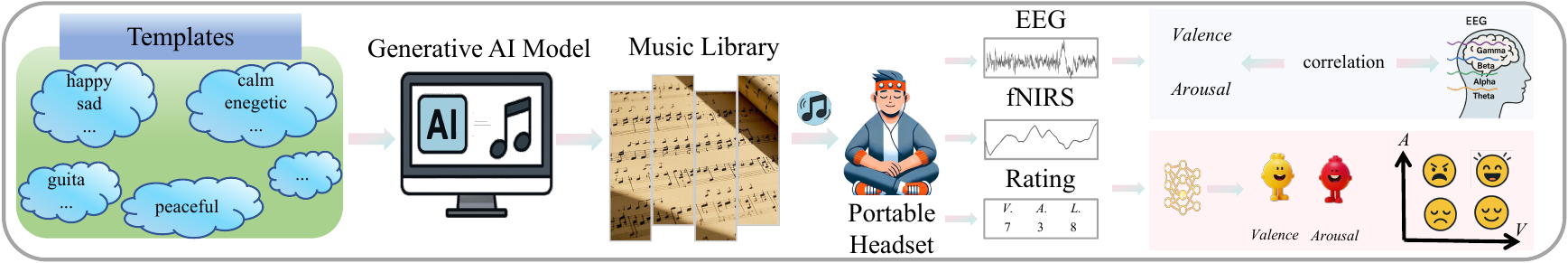}
		\caption{Overview of MEEtBrain framework.}
		\label{fig:teaser}
		\vspace{-0.1in}
	\end{figure*}
	
	Music-induced emotions are often accompanied by measurable physiological changes, making physiological signals a valuable resource for emotion recognition. Several modalities have been explored to assess emotional responses, such as heart rate, skin conductance, and respiratory patterns. It was 
	found that rhythmic features of music, such as tempo and articulation, influence these physiological signals, driving changes such as increased heart rate and faster breathing \cite{hu2018relationships, hsu2017automatic}. Techniques such as EEG, functional near-infrared spectroscopy (fNIRS), and positron emission tomography (PET) have also been employed to measure music-induced emotions.
	Notably, pleasant music has been shown to enhance prefrontal cortex activity while inhibiting the amygdala, facilitating emotional regulation \cite{moore2013systematic, lin2010eeg}.
	Recent advances in multi-modal frameworks integrate various physiological signals, such as EEG and fNIRS, to provide a comprehensive understanding of emotional responses to music  \cite{qiu2022multi, li2024eeg}.
	
	\subsection{Emotion Recognition Using Physiological Signals}
	
	EEG has been extensively studied as a tool for emotion recognition owing to its ability to capture real-time brain activity.
	Early methods relied on manually extracted features, such as power spectral density and coherence, combined with threshold-based rules for classification \cite{liu2010real}. In comparison, machine learning techniques, such as Linear Discriminant Analysis (LDA) \cite{wang2014emotional}, Support Vector Machines (SVM) \cite{lan2016real, wang2014emotional, lin2010eeg}, and K-Nearest Neighbor (KNN) \cite{mohammadi2017wavelet}, provide more sophisticated, data-driven approaches to pattern recognition. However, they still rely heavily on feature engineering. With the rapid development of deep learning techniques, models such as Convolutional Neural Networks (CNNs), Recurrent Neural Networks (RNNs), and Transformers have shown remarkable improvements in both classification accuracy and robustness \cite{zheng2015investigating, li2018hierarchical, wei2023tc, wang2025global, mathumitha2025emotion, cheng2025conditional}. For example, \citet{atkinson2016improving} employed kernel classifiers combined with a feature selection method to perform binary classification on the DEAP dataset \cite{koelstra2011deap}.
	
	While some studies have investigated the role of various physiological signals in emotion recognition and regulation \cite{vazquez2022emotion, kim2008emotion}, they seldom emphasize practical usability. For instance, \citet{kim2008emotion} employed a four-channel biosensor system to measure electromyogram (EMG), electrocardiogram (ECG), skin conductivity, and respiration changes for emotion recognition. However, the complexity of setup, and vulnerability to various interferences considerably limit the practical applicability of such systems in real-world environments.
	In our study, we attempt to strike a balance between device portability and accuracy in emotion recognition, using EEG-fNIRS multimodal data to achieve the highest accuracy while ensuring applicability to daily life.

	\section{MEEtBrain Framework}
	\subsection{Overview}
	We propose a novel framework, named MEEtBrain, to collect multimodal brain signals induced by AI-generated music for the analysis of human affective states, shown in Figure \ref{fig:teaser}. First, we construct an AI-generated music library as emotion stimuli, where music clips are automatically generated on a large scale by a generative AI model, addressing the copyright issues and overcoming the limitation of the music clips scale. For the automatic generation, we design prompt templates following the Russell's \textit{Valence-Arousal} circumplex \cite{russell1980circumplex} and considering personalized interests in music. We build different prompt sentences as input to the AI generative model for each emotion state, and the generated music clips are labeled with the corresponding emotion state. We then evaluate the obtained music clips by recruiting volunteers to rate, ensuring the matching between each music clip and its labeled emotion state. Sequentially, we design a paradigm for multimodal brain signals collection, consisting of two sessions and four blocks in each session. Each subject listens to different music clips and EEG and fNIRS are collected at the same time. In particular, each recruited subject is required to rate the music clip at the end of the collection procedure, including the valence, arousal, and liking. Based on the collected data, we conduct comprehensive analysis of emotion states, and even recognize emotion states from the signals.
	
	\subsection{AI-generated music library construction}
	\label{subsec:music-gen}
	
	\subsubsection{\textbf{Valence-Arousal music library design.}}
	
	Inspired by affective neuroscience studies\cite{koelstra2011deap, russell1980circumplex}, we construct a music library based on Russell’s \textit{Valence}-\textit{Arousal} model, which represents emotions in a 2D space. The \textit{valence} axis reflects emotional positivity, from unpleasant (e.g., sadness) to pleasant (e.g., joy), while the \textit{arousal} axis indicates emotional intensity, from low (e.g., calm) to high (e.g., excitement). Each music clip is assumed to evoke a specific emotional state. Following a protocol similar to that of DEAP\cite{koelstra2011deap}, we select clips representing four emotion types: \textit{HAHV}, \textit{HALV}, \textit{LAHV}, and \textit{LALV}.
	
	\begin{figure*}
		\centering
		\begin{minipage}{0.26\textwidth}
			\centering
			\includegraphics[height=4.2cm, width=\textwidth]{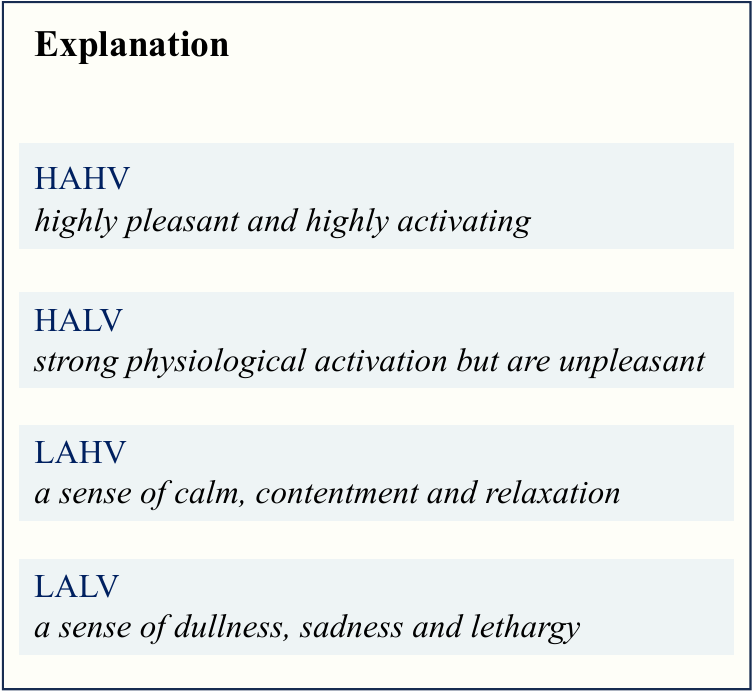}
			\subcaption{Four emotion categories.}
		\end{minipage}
		\begin{minipage}{0.26\textwidth}
			\centering
			\includegraphics[height=4.2cm, width=\textwidth]{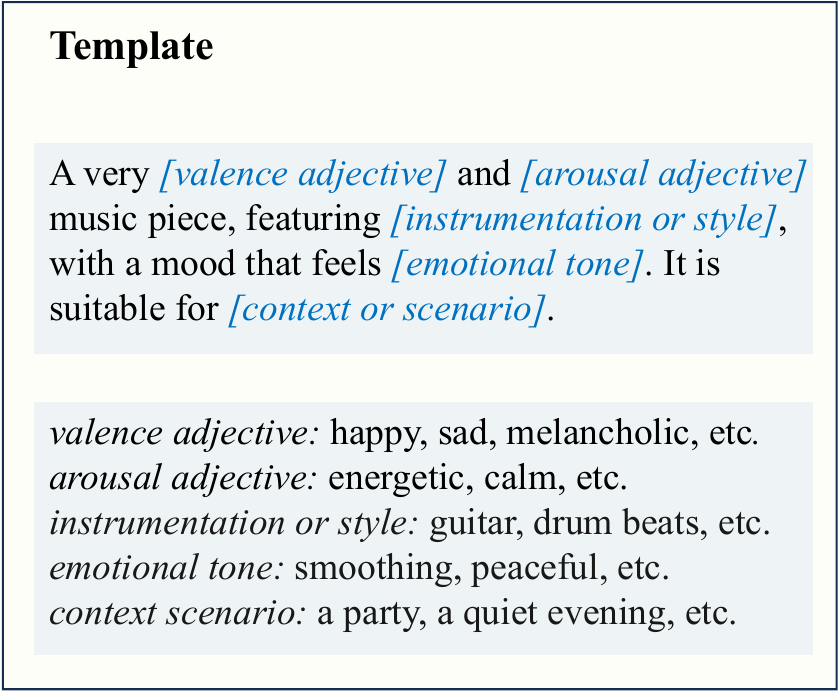}
			\subcaption{Template design.}
		\end{minipage}
		\begin{minipage}{0.26\textwidth}
			\centering
			\includegraphics[height=4.2cm, width=\textwidth]{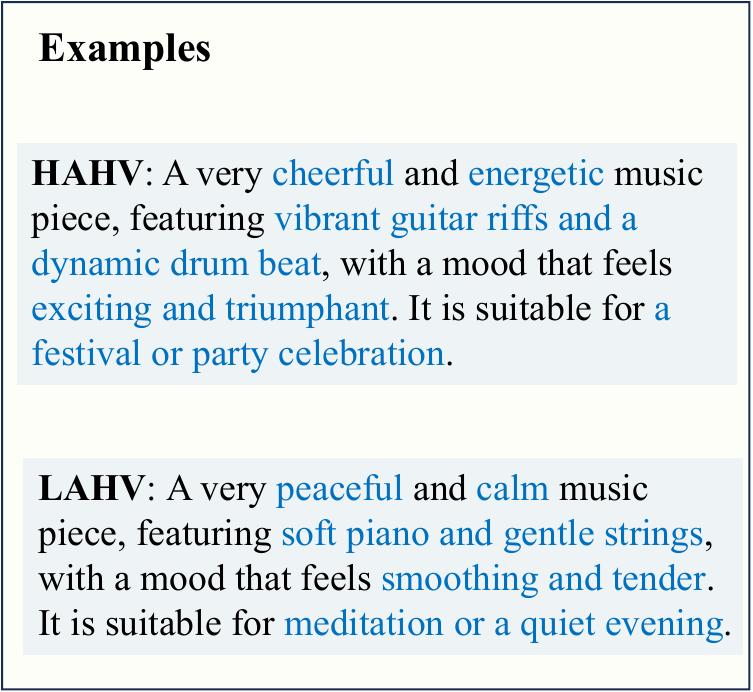}
			\subcaption{Examples of template.}
		\end{minipage}
		\caption{Explanation of Emotion Categories and Template Design. In figure (a), we provide a straightforward explanation of the four categories of emotions. In (b), we present our template along with several candidate words. In (c), we show two examples of template.}
		\label{fig:music-generation}
		\vspace{-0.1in}
	\end{figure*}
	\subsubsection{\textbf{Automatic music generation.}} To evoke emotions using music clips, we first need to create some music clips. Considering the copyright issues and limitation of music clip scale in existing studies, we try to automatically generate music clips taking advantage of the AIGC techniques, which can generate content based on the input keywords or requirements (prompt). In this way, the music clips can be automatically generated on a large scale. Here, we adopt  MUSICGEN model \cite{copet2023simple},  which is an autoregressive, transformer-based model for controllable music generation. It generates music clips considering both textual descriptions and melodic features. In other words, one prompt sentence is input, and MUSICGEN generates the required music clip. Intuitively, the input prompt is quite important for the music clips generated by MUSICGEN. Therefore, we devise a prompt template, making the generated music clips effectively evoke emotions. 
	
	Figure \ref{fig:music-generation} shows our prompt template: a single sentence describing music from three aspects—\textbf{specific emotions}, \textbf{instrument styles}, and \textbf{contextual scenarios}. Placeholders are filled with descriptive elements to customize prompts. The \textit{valence adjective} indicates emotional polarity (e.g., happy, sad), while the \textit{arousal adjective} reflects intensity (e.g., energetic, calm). The \textit{instrumentation or style} specifies instruments or genre (e.g., guitar, piano, drum beats), considering personal preference. The \textit{emotional tone} conveys the overall mood (e.g., soothing, exciting), and the \textit{context or scenario} suggests appropriate use cases (e.g., party, quiet evening). Prompts cover four emotional states: \textit{HAHV}, \textit{HALV}, \textit{LAHV}, and \textit{LALV}. We generate hundreds of music clips using iterative prompt variations via MUSICGEN, labeling each clip based on its prompt. In total, 236 clips are created, evenly distributed across the four emotion types.
	
	\subsubsection{\textbf{Music clips screening.}}
	
	\begin{figure}
		\centering
		\includegraphics[scale=0.45]{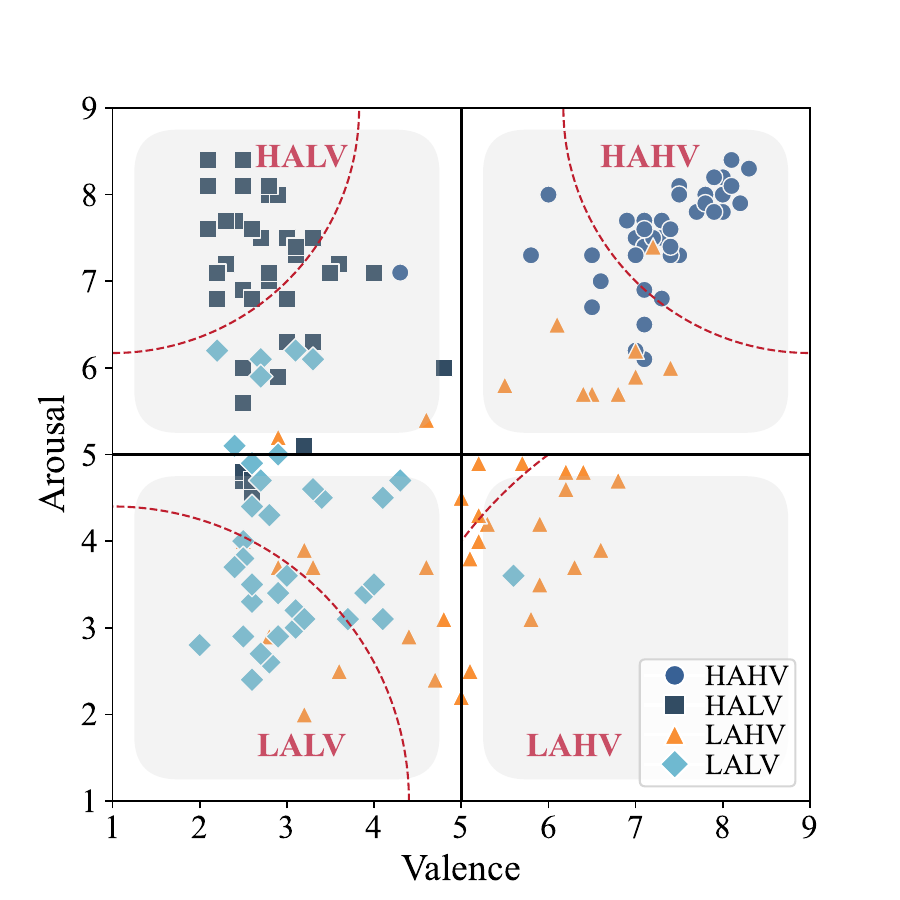}
		\caption{The rating score of each generated music clip given by the volunteers. Each point represents a music clip, with its coordinates corresponding to the scores assigned by the volunteers. The legend displays the labels initially assigned to the music clips, while the red dashed line delineates the final decision boundary for the music clips.}
		\label{fig:music-scores}
		\vspace{-0.2in}
	\end{figure}
	
	In order to guarantee the matching between the music clips and the labeled emotion state, we perform a music clip screening process. We first exclude music clips with technical flaws (e.g., abrupt noise, extended silent intervals), and there are 157 ones retained. Then, we recruit volunteers to rate the generated music clips by scoring each clip on a scale from 1 to 9 along two dimensions: \textit{Valence} and \textit{Arousal}. A \textit{Valence} score of 1 denotes "painful", whereas a score of 9 represents "pleasant". Similarly, an \textit{Arousal} score of 1 indicates low arousal, while a score of 9 signifies high arousal. Figure \ref{fig:music-scores} illustrates the rating outcomes from the 10 evaluators for the 157 music clips.
	As we expected, the majority of the music clips reliably evoked the target emotions, and the evaluators' scores closely align with the labels of the music clips. 
	We further select the music clips that significantly align with the intended emotional states according to the distribution shown in Figure \ref{fig:music-scores}. As shown, different emotion states are concentrated in different regions, and so we adopt different strategies to select music chips for the various emotional categories. For the \textit{HAHV} category, we select the clips that satisfy the condition $\sqrt{(v-9)^2 + (a-9)^2} \leq 2\sqrt{2}$, where $(v,a)$ represents the \textit{Valence-Arousal} scores of each clip. For instance, the emotional responses are associated with \textit{HAHV} when both \textit{Valence} and \textit{Arousal} scores exceed 7. Similarly, for the \textit{HALV} and \textit{LALV} categories, clips are retained if they meet the conditions $\sqrt{(v-1)^2+(a-9)^2} \leq 2\sqrt{2}$ and $\sqrt{(v-1)^2 + (a-1)^2} \leq 3.4$, respectively. 
	For the \textit{LAHV} category where the clips are not so concentrated, to keep the data balance, we restrict our selection to clips that satisfy the conditions $\sqrt{(v-9)^2+(a-1)^2} \leq 5$, in conjunction with $v \geq 5$ and $a \leq 5$.
	Finally, we obtain a total of 101 music clips, composed of 31 clips for the \textit{HAHV} category, 23 for \textit{HALV}, 21 for \textit{LAHV}, and 26 for \textit{LALV}.

	\subsection{Multimodal brain signals collection}
	
	\begin{figure*}
		\centering
		\includegraphics[scale=0.45]{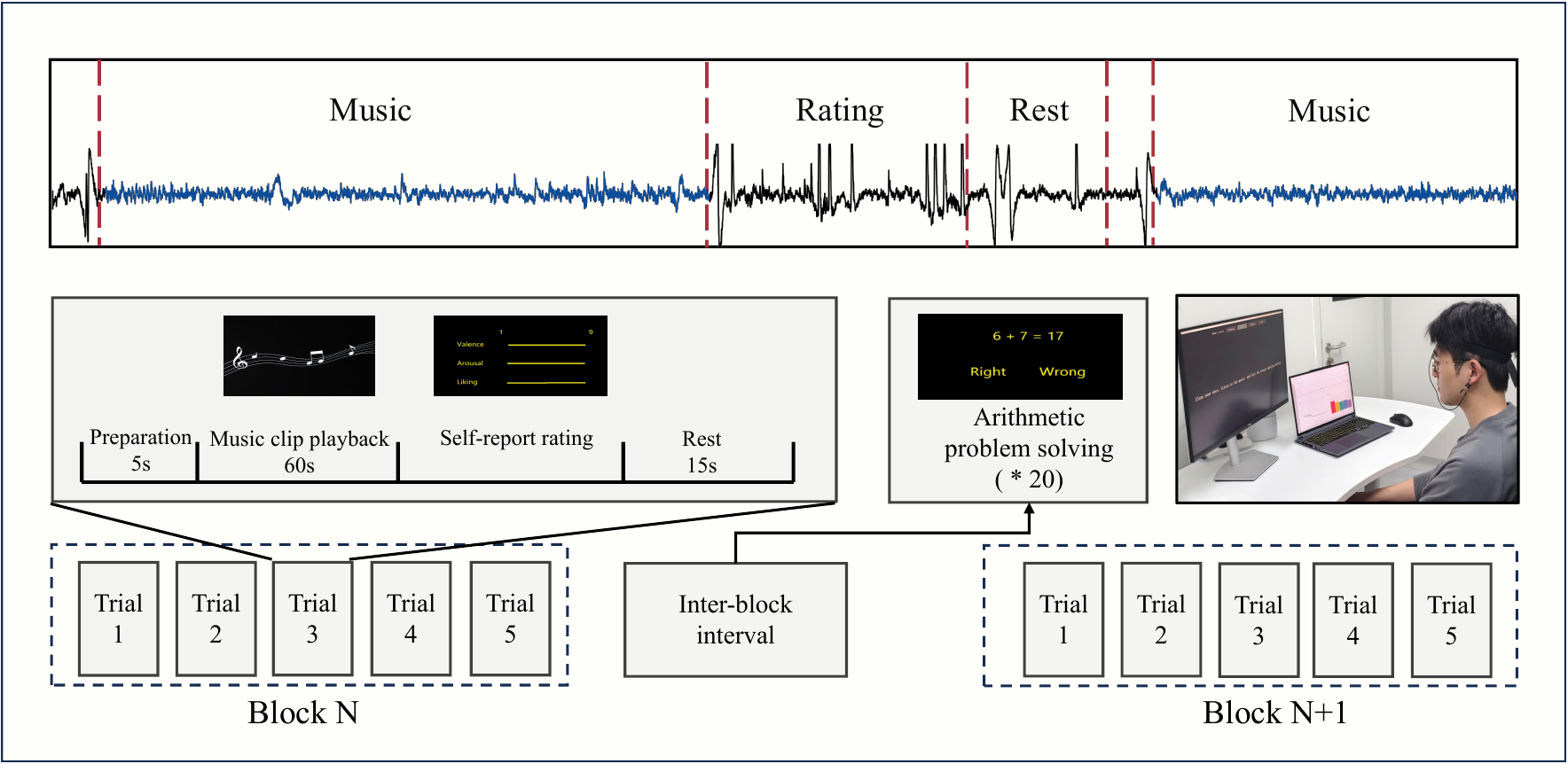}
		\caption{Multimodal brain signals collection paradigm.}
		\label{fig:paradigm}
		\vspace{-0.1in}
	\end{figure*}
	
	\subsubsection{\textbf{Collection paradigm.}}
	On the basis of the screened music clips, we collect multimodal brain signals of EEG and fNIRS. We first design a collection paradigm inspired by the collection of FACED dataset \cite{chen2023large}, shown in Figure \ref{fig:paradigm}. There are two sessions involved, and four blocks in each session. Each block consists of five trials, and each trial lasts around 90 seconds. Prior to the start of each trial, there is a 5-second preparation period when participants are instructed to close their eyes and adjust their state of mind. Then, the music is played in 60 seconds, during which the participants are required to keep their eyes closed, remain still, and avoid head or eye movements to better focus on the music and engage with its emotional nuances. \textbf{After the music concludes, the participants rate the \textit{Valence}, \textit{Arousal}, and \textit{Liking} of the piece}. The \textit{Liking} should not be confused with \textit{Valence}. It explores one's tastes or interests in the music they listened to rather than their emotional responses. For example, some individuals may enjoy music that makes them feel sad \cite{koelstra2011deap}. At the end of each complete trial, participants are allowed to rest for 15 seconds. To minimize the potential influence of emotional alternation, the same type of emotional music is used for all the five trials within a single block. Additionally, participants are required to solve a few simple arithmetic problems between blocks to maintain their focus. Under this paradigm, each participant completes two sessions in 80 minutes.
	
	\subsubsection{\textbf{Portable and multimodal collection device.}}
\begin{figure}
		\centering
		\includegraphics[scale=0.4]{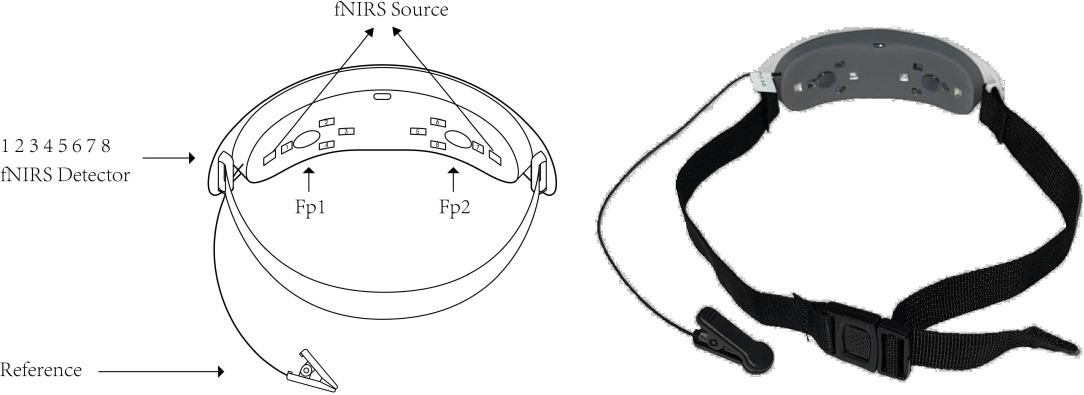}
		\caption{Headband device for collecting EEG and fNIRS}
		\label{fig:headset}
		\vspace{-0.2in}
	\end{figure}

	To promote the accessibility and friendliness of the framework usage, we develop a wireless portable EEG headband for multimodal brain signals acquisition, shown in Figure \ref{fig:headset}. It integrates dry sensor technology to simultaneously record EEG and fNIRS signals, with high signal quality and motion immunity for real-world applications. It has two notable advantages: 1) \textbf{Portable and user-friendly}: compared to traditional EEG caps, the device is designed in headband-style, making it easy to wear in forehead. It uses dry electrodes, eliminating the need for conductive gel, making it convenient and user-friendly. 2) \textbf{Multimodal integrated}: it can simultaneously collect EEG and fNIRS signals. The two different signal types reflect brain activities from complementary perspectives. Its signal acquisition performance has been evaluated in \cite{li2024hybrid}. 
	
	We obtain high-quality data with its capability to capture event-related potentials and hemodynamic responses in the prefrontal area. Specifically, the EEG signals are collected at a rate of 250 HZ from Fp1 and Fp2 electrodes referenced to the left earlobe A1 according to the international 10–20 system. Simultaneously, fNIRS signals are recorded, capturing hemodynamics from the bilateral frontal cortex with 8 optodes. The fNIRS data is sampled at 25 Hz using two wavelengths: infrared (850 nm) and near-infrared (735 nm). Its modular design minimizes spatial interference and signal crosstalk, while an integrated EEG pre-amplifier enhances signal quality by reducing noise and distortion. Advanced filtering and motion correction techniques further refine the fNIRS signals, and a high source-switching frequency improves crosstalk suppression.

	\begin{table*}
		\centering
		\renewcommand{\arraystretch}{1.5} 
		\caption{A summary of key features of our dataset.}
		\label{tab:summary}
		\begin{tabular}{lll}
			\hline
			\multicolumn{3}{c}{\textbf{Key features of the dataset}} \\
			\hline
			Number of participants & 44 (24 newly recruited) & mean age: 25.75 years, native Chinese \\
			\hline
			Emotion categories & high/low \textit{Arousal/Valence} & \textit{HAHV, HALV}, \textit{LAHV, LAHV} \\
			\hline
			Stimuli & 101 music clips & \makecell{31 clips for the \textit{HAHV}, 23 for \textit{HALV}, 21 for \textit{LAHV}, \\and 26 for \textit{LALV}, 30 seconds each clip}\\
			\hline
			Signals duration & 1760 minutes & 44 (participants) * 8 (blocks) * 5 (trials) * 60 (seconds)\\
			\hline
			EEG signals & 2 channels & Fp1, Fp2 according to the international 10-20 system \\
			\hline
			fNIRS signals & 8 channels & all placed in the prefrontal area \\
			\hline
			Self-reporting ratings & 3 items & \textit{Arousal}, \textit{Valence}, \textit{Liking} (scale of 1-9)\\
			\bottomrule
			
		\end{tabular}
	\end{table*}
	
	\section{Experiment}
	
	\subsection{Data collection and overview}
	\subsubsection{\textbf{Data collection.}}On the basis of the MEEtBrain framework, we recruited participants for data collection. The volunteers with a history of psychological or psychiatric disorders are excluded. There were 20 participants for the experiment (44 participants in the latest dataset). To ensure that the data collection process is not affected by external environmental interference, the participants completed the data collection in a shielded room. They sat wearing the EEG headband, positioned 60 cm away from the monitor screen. To protect user privacy, each participant was anonymized and assigned a unique identifier (\textit{S01}, \textit{S02}, ..., \textit{S44}) throughout data analysis. Each music clip is played twice to ensure it effectively evokes emotions. The study was approved by the ethics committee of the Hangzhou Seventh People's Hospital (Approval No. 2024-035).
	
	\subsubsection{\textbf{Data labels.}}Participants provided self-reporting ratings for three dimensions: \textit{Arousal}, \textit{Valence}, and \textit{Liking}. It is worth noting that, \textbf{for each EEG and fNIRS recording, its emotion label is from the rating scores of valance and arousal given by the corresponding participant. }Specifically, the arousal or valence label of the data is based on a 5-point threshold, where a score above 5 is high and a score below 5 is low. In particular, when the score is 5, the label of the data is marked as the arousal or valence of the music. For example, for one recording, if the subjective rating score of arousal and valance are 3 and 7, respectively, its emotion label is LAHV.
	
	\subsubsection{\textbf{Data overview.}}Table \ref{tab:summary} provides a concise summary of the key features of the dataset used in the study.
	The dataset was collected from 44 native Chinese participants (24 newly recruited), age ranging from 22 to 38 years, with a mean age of 25.75 years. The emotional states are categorized into four distinct categories based on arousal and valence levels: low \textit{Arousal} low \textit{Valence}, low \textit{Arousal} high \textit{Valence}, high \textit{Arousal} low \textit{Valence}, and high \textit{Arousal} high \textit{Valence}. We used 101 music clips to evoke emotions, 31 clips for the \textit{HAHV} category, 23 for \textit{HALV}, 21 for \textit{LAHV}, and 26 for \textit{LALV}. To facilitate evaluation, we have randomly chosen 3 AI-generated music clips for each emotional state and included them in the supplementary materials. We collected 44 recordings of EEG and fNIRS simultaneously, each one lasts 2400 seconds and there are 1760 minutes in total in duration. The EEG recordings are from 2 channels (FP1, FP2) and fNIRS recordings are from 8 channels (placed in the prefrontal area). \textbf{The dataset is available at} \url{https://zju-bmi-lab.github.io/ZBra}.
	
	\subsection{Data preprocessing}
	
	\subsubsection{\textbf{EEG preprocessing.}} We first performed preprocessing on EEG and fNIRS recordings, respectively. EEG was pre-processed through MNE-python toolbox \cite{gramfort2013meg}. We manually excluded the segments affected by device-related artifacts. Then, a bandpass filter (0.1–40 Hz) was applied. To ensure the alignment between each EEG segments and its emotional label of each music clip, the recordings were epoched from the 25th to 55th seconds post-stimulus onset, that are assumed to capture the critical phase of the trial. Finally, baseline correction was performed on each epoched segment.
	
	\subsubsection{\textbf{fNIRS preprocessing.}} fNIRS preprocessing was conducted using Python 3.12 and SciPy 1.15, following standardized pipelines. First, we converted raw optical signals to optical density changes, and a 0.5–4 Hz band-pass filter was applied to obtain the photoplethysmogram (\textbf{PPG}) signal, which reflects heart rate. Subsequently, concentration changes in oxyhemoglobin (\textbf{HbO}), deoxyhemoglobin (\textbf{HbR}), and total hemoglobin (\textbf{HbT}) were computed using the Modified Beer-Lambert Law \cite{bouguer1729essai, lambert1760photometria, izzetoglu2020effects}, a widely adopted method for converting light intensity measurements into hemodynamic metrics. Baseline correction was performed by subtracting the mean value of a 5-second pre-stimulus baseline from each channel's signal during the task, ensuring task-related activity was decoupled from resting-state variations.
	Finally, a 0.01–0.1 Hz band-pass filter was applied to remove low-frequency systemic artifacts, including cardiac pulsations ($\le$ 1 Hz), respiratory venous waves ($\le$ 0.2 Hz), and Mayer wave oscillations ($\le$ 0.1 Hz), which are known to confound fNIRS signals \cite{qiu2022multi}.
	
	\begin{figure}
		\centering
		\includegraphics[scale=0.33]{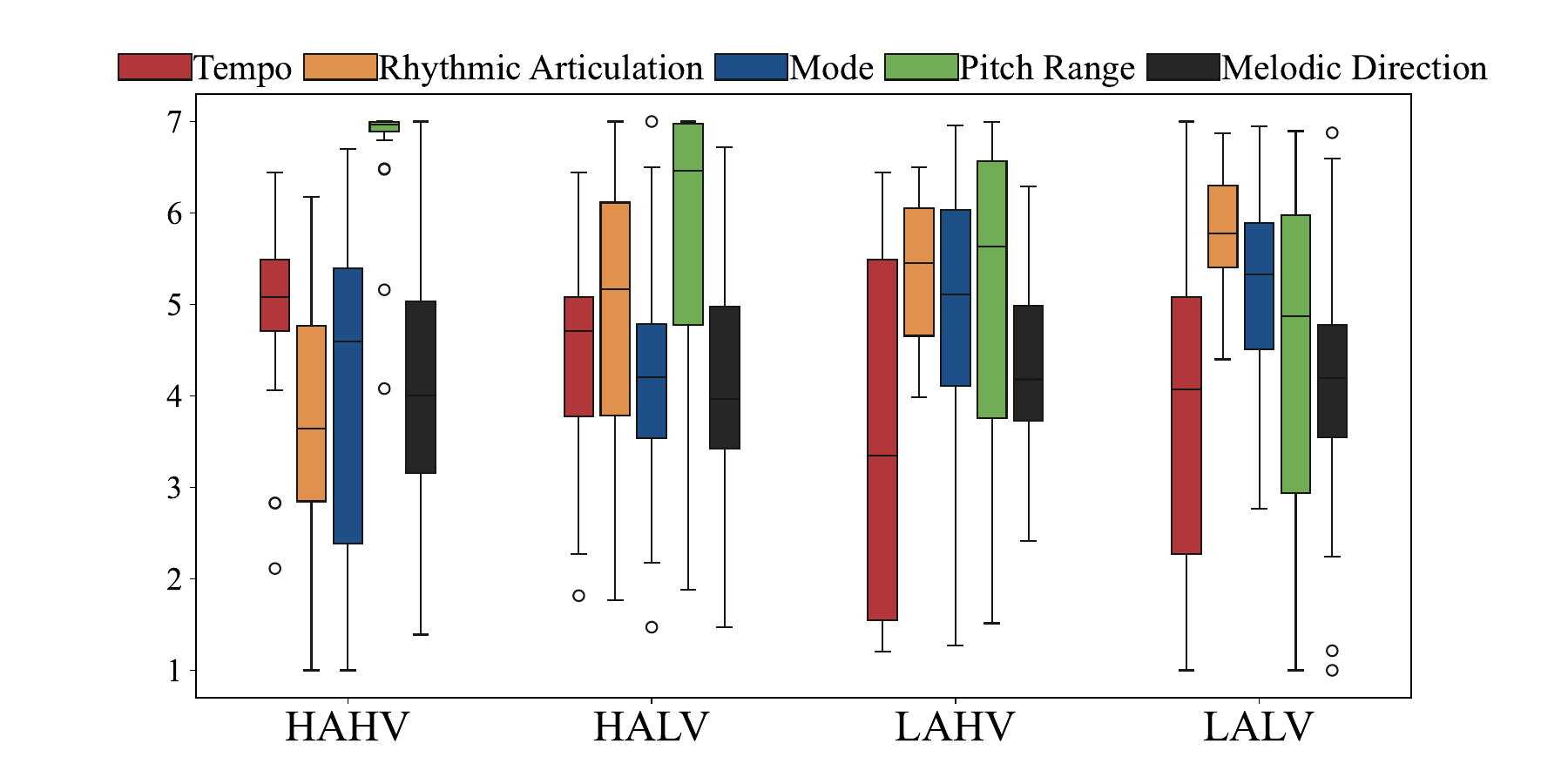}
		\caption{Structural features of music (linearly scaled to the range [1,7]) across four music groups.}
		\label{fig:music-structure}
		\vspace{-0.2in}
	\end{figure}
	
		%

	\section{Results and Analysis}
	
	\subsection{Validation of the AI-generated music clips}
	Previous studies show that mode and rhythmic articulation significantly affect \textit{Valence}, while tempo and rhythmic articulation influence \textit{Arousal}; melodic direction shows the weakest emotional correlation \cite{gomez2007relationships, juslin1997perceived}. To validate this on our generated music, we extracted the following structural features: 1) \textbf{Tempo}: which measures music speed in beats per minute (BPM) and ranges from slow to fast, estimated via beat-tracking \cite{ellis2007beat}; 2) \textbf{Rhythmic Articulation}, describing how notes are performed—ranging from staccato (short, detached) to legato (smooth, connected)—is derived from the inverse of the mean zero-crossing rate (ZCR); 3) \textbf{Mode}, which defines the tonal quality of a piece (e.g., minor for sadder tones and major for happier tones), is determined by comparing chroma features (pitch class distributions) to predefined major and minor templates using dot product similarity; 4) \textbf{Pitch Range}, referring to the span between the highest and lowest notes in a piece, is measured as the difference between the highest and lowest valid pitches; 5) \textbf{Melodic Direction}, indicating the overall pitch movement in a melody (ascending or descending), is analyzed by computing the ratio of ascending to descending intervals, with scores inversely mapped to emphasize descending trends. All the extracted metrics were further scaled to a range of [1,7] for consistency and analysis.
	
	The results are shown in the Figure \ref{fig:music-structure}. To assess group differences, we conducted a one-way ANOVA test on the structural features across the four emotion categories (\textit{HAHV}, \textit{HALV}, \textit{LAHV}, \textit{LALV}). As we can see, rhythmic articulation and pitch range exhibit a highly significant effect on the \textit{Arousal} dimension ($p$<0.00001). Mode (major/minor) and tempo showed statistically significant differences across the four groups ($p$<0.05). Melodic direction displayed no significant variation ($p$>0.5). \textbf{These findings are consistent with the conclusions from prior work \cite{gomez2007relationships, juslin1997perceived}, indicating the effectiveness of our generated music clips in evoking emotions.}
	
	\begin{figure}
		\centering
		\includegraphics[scale=0.3]{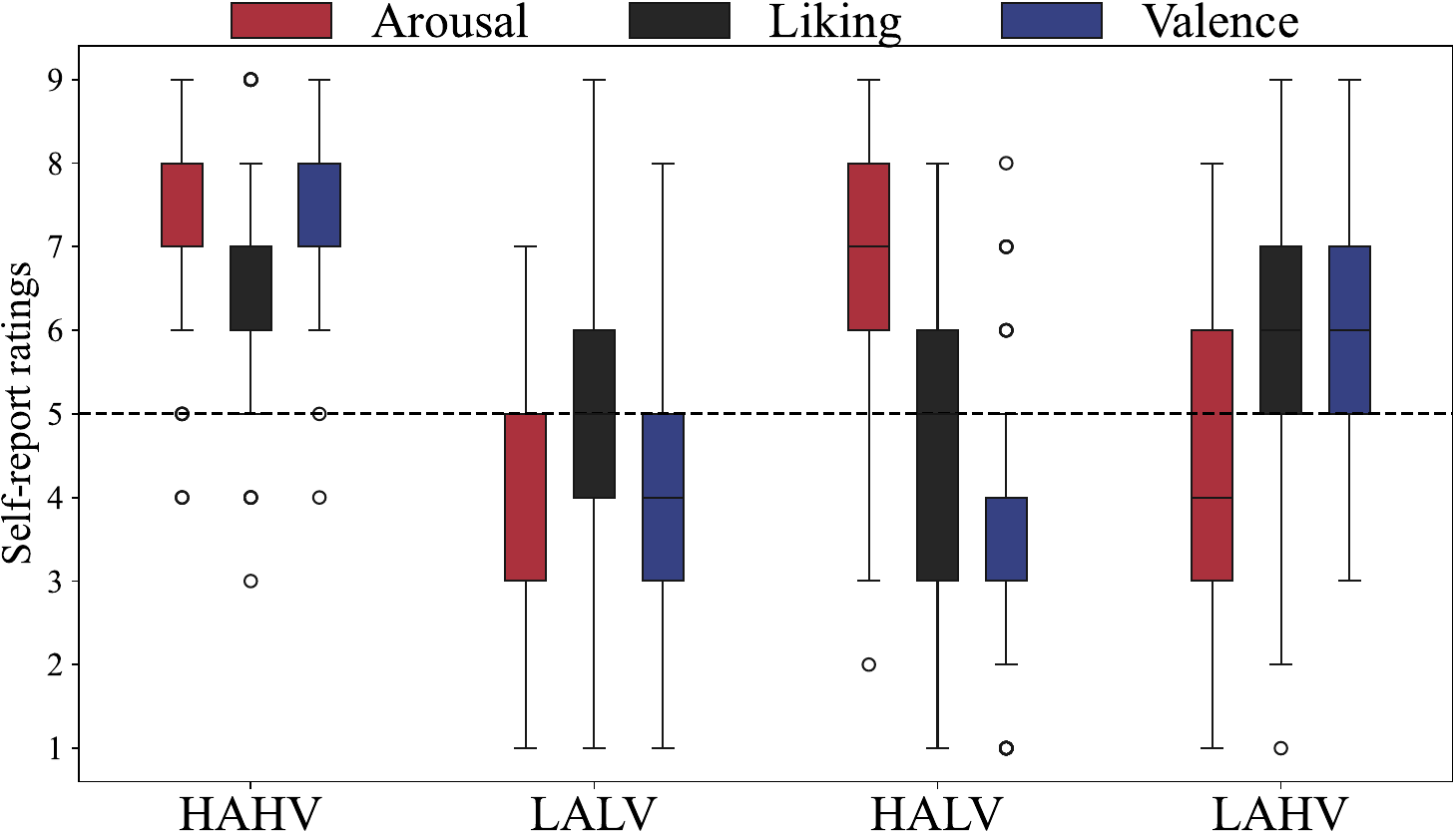}
		\caption{Distribution of self-report rating scores across music groups during the collection experiment.}
		\label{fig:rating-distribution}
		\vspace{-0.1in}
	\end{figure}
	
	\subsection{Evaluation of the emotion-inducing effect}
	
	We further evaluated the emotion-inducing effect of our generated music, by comparing the rating scores given by the participants and the initial labels of the music clips. We conducted a statistical analysis of \textit{Valence}/\textit{Arousal} scores of the participants in each emotion groups, shown in Figure \ref{fig:rating-distribution}. As we can see, \textbf{most rating scores given by the participants are aligned with the emotion of each group}. It further indicates that \textbf{our generated music can effectively evoke the corresponding emotion states}. When the valence score is relatively high, the liking score is also higher, indicating \textbf{most participants prefer to the music that can evoke high valance} (the valence score and liking score are strongly positively correlated, shown in Table \ref{tab:items-correlations}). There are few outliers for each group, which may be due to the individual differences in music. 
	
	We examined correlation of \textit{Liking}, \textit{Arousal}, and \textit{Valence} scores, by conducting a Pearson correlation analysis shown in Table \ref{tab:items-correlations}. As expected, we found \textbf{a significant and strongly positive correlation between between \textit{Valence} and \textit{Liking}}. Most individuals prefer to music that evokes positive emotions such as happiness and joy, although some may have a preference to sad music. 
	Besides, \textit{Arousal} and \textit{Valence} are not independent, exhibiting a relatively low correlation. This finding is consistent with the results found in DEAP dataset \cite{koelstra2011deap}.
	
	\begin{table}
		\centering
		\caption{The correlations among the rating scores of \textit{Arousal}, \textit{Valence}, \textit{Liking}. ($*$:  p-value < 0.001)}
		\label{tab:items-correlations}
		\begin{tabular}{cccc}
			\toprule
			Rating items & \textit{Valence} & \textit{Arousal} & \textit{Liking} \\
			\midrule
			\textit{Valence} & 1.0 & 0.212* & 0.602* \\
			\textit{Arousal} & 0.212* & 1.0 & 0.016 \\
			\textit{Liking} & 0.602* & 0.016 & 1.0 \\
			\bottomrule 
		\end{tabular}
	\end{table}
	
\begin{figure}
		\centering
		\includegraphics[scale=0.3]{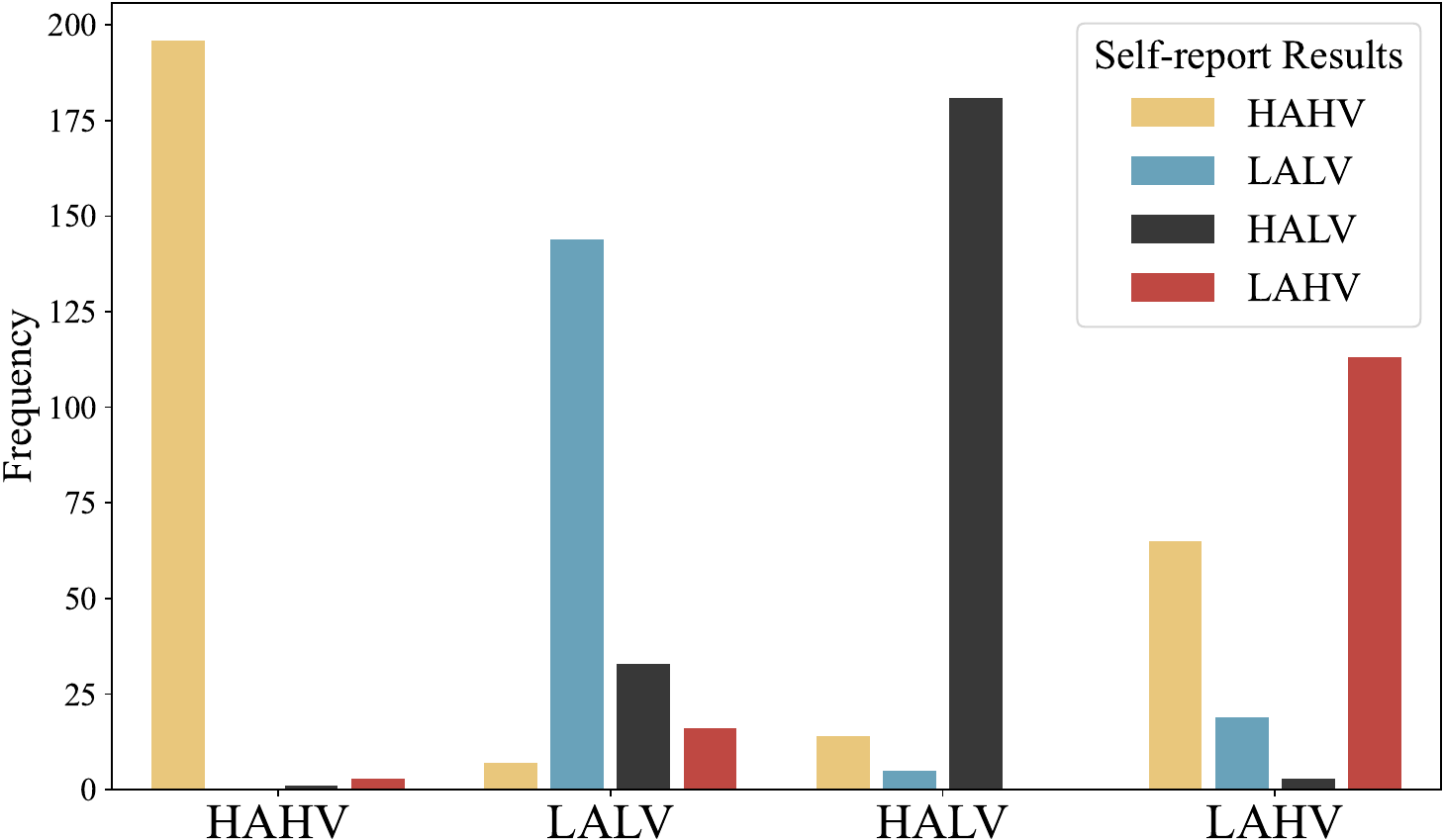}
		\caption{Histogram of participants with rating scores for each music category.}
		\label{fig:validation-elicit}
		\vspace{-0.1in}
	\end{figure}

	For each recording, we also plotted a histogram, where the x-axis is the emotion groups, and the y-axis is the frequency of music clips with the corresponding emotion label, shown in Figure \ref{fig:validation-elicit}. It can be seen, in each emotion group, \textbf{each type of music successfully induces a high proportion of the targeted emotions}. It further proves that our generated music can effectively evoke the corresponding emotion. Besides, some \textit{LAHV} stimuli induced higher arousal than expected, shown in both Figure \ref{fig:music-scores} and \ref{fig:validation-elicit}.
	This is consistent with the well-validated ratings for the International Affective Picture System (IAPS) \cite{lang1997international} and the International Affective Digital Sounds system (IADS) \cite{bradley1999international}, as well as in the study by DEAP \citet{koelstra2011deap}. This suggests a general difficulty in eliciting emotions by high valence but low arousal.

\begin{table*}[htbp]
	\centering
	\caption{Cross-validation results of emotion recognition. Note: PPG refers to photoplethysmogram; Hb refers to HbO/HbR/HbT.}
	\label{tab:cross-validation}
	\vspace{-0.08in}
	
	\begin{subtable}{\textwidth}
		\centering
		\caption{Cross-subject results}
		\begin{tabular}{lcccc}  
			\toprule
			\makecell[c]{Modality} & \multicolumn{2}{c}{\textit{Valence}} & \multicolumn{2}{c}{\textit{Arousal}} \\
			\cmidrule(lr){2-3} \cmidrule(lr){4-5}
			& ACC & MF1 & ACC & MF1 \\
			\midrule
			EEG & 0.631±0.070 & 0.622±0.073 & 0.688±0.062 & 0.634±0.044 \\
			PPG & 0.646±0.071 & 0.593±0.111 & 0.671±0.092 & 0.562±0.114 \\
			Hb & 0.625±0.054 & 0.564±0.097 & 0.665±0.119 & 0.518±0.144 \\
			EEG+PPG & 0.658±0.064 & 0.623±0.080 & 0.697±0.074 & 0.643±0.057 \\
			EEG+Hb & 0.649±0.078 & 0.622±0.092 & 0.704±0.081 & 0.635±0.069 \\
			\textbf{EEG+PPG+Hb} & \textbf{0.663±0.064} & \textbf{0.636±0.078} & \textbf{0.719±0.077} & \textbf{0.657±0.065} \\
			\bottomrule
		\end{tabular}
	\end{subtable}
	
	\vspace{1.5em}
	
	\begin{subtable}{\textwidth}
		\centering
		\caption{Intra-subject results}
		\begin{tabular}{lcccc}
			\toprule
			\makecell[c]{Modality} & \multicolumn{2}{c}{\textit{Valence}} & \multicolumn{2}{c}{\textit{Arousal}} \\
			\cmidrule(lr){2-3} \cmidrule(lr){4-5}
			& ACC & MF1 & ACC & MF1 \\
			\midrule
			EEG & 0.640±0.077 & 0.633±0.082 & 0.709±0.067 & 0.649±0.076 \\
			PPG & 0.646±0.059 & 0.632±0.063 & 0.685±0.076 & 0.610±0.041 \\
			Hb & 0.643+0.056 & 0.633±0.061 & 0.670±0.074 & 0.572±0.061 \\
			EEG+PPG & 0.656±0.070 & 0.653±0.070 & \textbf{0.713±0.078} & 0.653±0.095 \\
			EEG+Hb & 0.659±0.084 & 0.654±0.088 & \textbf{0.713±0.076} & \textbf{0.659±0.085} \\
			\textbf{EEG+PPG+Hb} & \textbf{0.679+0.066} & \textbf{0.673±0.069} & 0.710±0.065 & 0.657±0.076 \\
			\bottomrule
		\end{tabular}
	\end{subtable}
	
	\vspace{-0.1in}
\end{table*}

	\begin{table}
		\centering
		\caption{Statistical significance of power spectral differences across frequency bands of EEG in four emotion groups.}
		\vspace{-0.05in}
		\label{tab:significant}
		\begin{tabular}{cccccc}
			\toprule
			\textbf{band} & delta & theta & alpha & beta & gamma \\
			\midrule
			\textbf{p-value} & 0.705 & 0.348 & 0.068 & 5.483e-05 & 0.321 \\
			\bottomrule
		\end{tabular}
		\vspace{-0.15in}
	\end{table}
	
	\subsection{Brain signals and emotions}
	\subsubsection{\textbf{EEG and emotion.}} To identify the neural correlation of emotional experience, we analyze the relationship between the relative EEG spectral powers and self-reported ratings. The relative spectral power refers to the ratio between the sum of the spectral powers in the frequency band of interest (delta, theta, alpha, beta, and gamma) and the sum of full-band spectral power \cite{chen2023large}. Here, we focused on the final 30 seconds of the 60-second music period, and divided it into multiple 3-second epochs. For each epoch, we calculated the sum of spectral power within the five frequency bands, and calculated the total spectral power across all the frequencies, averaging these values across two EEG channels. Finally, we computed the mean relative spectral power for all the epochs within the 30-second window for each trial. To determine whether significant differences existed in mean relative spectral power among different emotion groups, we performed a one-way ANOVA test, shown in Table \ref{tab:significant}.  \textbf{There are significant differences in the mean relative power of the \textit{beta frequency bands }across the music groups ($p$<0.001).} Subsequently, we conducted post hoc tests using Tukey’s Honestly Significant Difference (HSD) test for the beta frequency bands \cite{abdi2010tukey}. There are significant differences in beta band power between the \textit{LAHV} group and the other three groups.
	
	\subsubsection{\textbf{fNIRS and emotion.}} 
	
	We also explored how fNIRS signals reflect emotion. To this end, for each 60-second epoch, we computed the mean and variance of HbO, HbR, and HbT concentrations using data from the last 30 seconds across all the eight channels. This process yielded a 48-dimensional feature set. Subsequently, we employed Pearson’s correlation analysis to investigate the relationship between these features and the corresponding emotion labels. Some certain features exhibit significant correlations with the emotion labels, such as the variance of HbR concentration in channel 3 ($p$<0.01). Notably, \textbf{all the significant correlations are observed between the variance metrics and the emotion labels, suggesting that the dynamic changes in hemoglobin concentration may encode emotion-related information}. These findings highlight the potential of fNIRS-based measures to capture neurophysiological signals related to emotional states.
	
	\subsection{Emotion recognition}
	
	We tried to recognize emotion using the collected brain signals from only the first 20 participants. Here, the emotion recognition task refers to binary classification of valence and arousal, respectively. We explored the predictive ability of the multimodal brain signals. For the multimodal signals: EEG, fNIRS-derived hemodynamic signals (HbO/HbR/HbT) and photoplethysmography (PPG), we evaluated the classification performance across six modality combinations: (1) EEG-only, (2) PPG-only, (3) HbO/HbR/HbT-only, (4) EEG+PPG, (5) EEG+HbO/HbR/HbT, (6) EEG+PPG+HbO/HbR/HbT. For classification, we implemented a novel deep neural network architecture, extending the Conformer model \cite{song2022eeg} with a dedicated fNIRS-specific branch for signal fusion. We tested the performance under both cross-subject and intra-subject paradigms. For cross-subject classification, a leave-one-subject-out strategy was adopted, and for intra-subject classification, we used the 10-fold cross validation strategry. Mean accuracy (ACC) and macro-averaged F1 score (MF1) were used for performance measurement.
	
	As we can see from Table \ref{tab:cross-validation}, \textbf{the combination of all the modalities performs the best across most of the tasks}. Integrating multiple modalities significantly improves classification performance for both \textit{Valence} and \textit{Arousal}, compared to using single modality. Notably,\textbf{ PPG is the most powerful for \textit{Valence} recognition in ACC among the multiple modalities, and EEG performs the best in \textit{Arousal} classification.} Meanwhile, PPG performs better in classifying \textit{Arousal} than \textit{Valence}, around 3\% higher in ACC. Individuals' heart rate (PPG) is likely to exhibit noticeable fluctuations when listening to fast-paced music that could affect arousal.
	Interestingly, \textbf{\textit{Arousal} classification performance is consistently better than that of \textit{Valence} in ACC and MF1 across all the tasks}. This suggests that \textit{Arousal} is more reliably detectable using the available physiological signals.

	\vspace{-0.15in}
	\section{Conclusion}
	
	In order to promote the accessibility of emotion regulation, we propose a portable and multimodal framework for analyzing emotions induced by AI-generated music and collecting EEG and fNIRS through a portable headband. We address three key limitations of existing studies: (1) restricted stimulus diversity, (2) single-modality of signals, and (3) lack of portability. AIGC techniques are introduced to automatically generate music stimuli based on our designed prompt templates. The music stimuli can be generated on a large scale, and independent of subjective selection. Then we use our portable headband to collect EEG and fNIRS signals when participants are listening to the music stimuli. We collected a real-world dataset of EEG and fNIRS from 20 individuals in the first recruitment, with a total duration of about 14 hours. Based on the dataset, the generated music clips were verified and the emotion-inducing effects were evaluated, indicating the generated music effectively evoke the intended emotions. We also successfully recognize emotions combining EEG and fNIRS. We are actively expanding our multimodal dataset (44 participants in the latest dataset) and make it publicly available to promote further research and practical applications. \textbf{The dataset is available at} \url{https://zju-bmi-lab.github.io/ZBra}.
	
	\section*{Acknowledgments}
	This work was supported by STI 2030 Major Projects (2021ZD0200400), National Natural Science Foundation of China  (No.62476240) and the Key Program of the Natural Science Foundation of Zhejiang Province, China (No. LZ24F020004). The corresponding author is Dr. Gang Pan.
	
	
	\bibliographystyle{ACM-Reference-Format}
	\bibliography{sample-base}
	
\end{document}